\documentclass{moriond}

\def\be{\begin{equation}}
\def\ee{\end{equation}}
\def\bea{\begin{eqnarray}}
\def\eea{\end{eqnarray}}

\newcommand{\Photo}{\includegraphics[height=35mm]{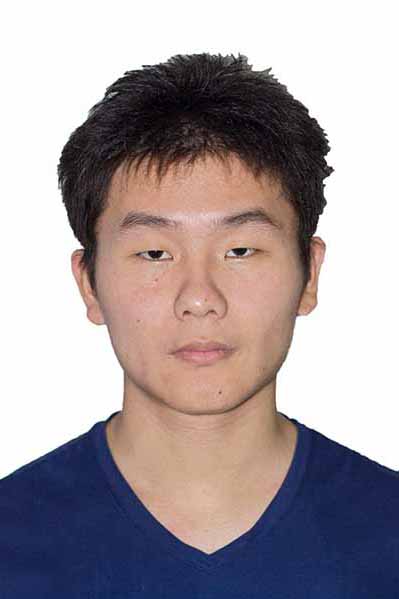}}

\begin{document}
\vspace*{4cm}
\title{SEARCH FOR A HEAVY RESONANCE DECAYING  TO A PAIR OF VECTOR BOSONS IN THE LEPTON PLUS MERGED JET FINAL STATE AT $\sqrt{s}$ = 13TeV}

\author{HUANG HUANG}

\address{On behalf of the CMS collaboration \\
State Key Laboratory of nuclear Physics and Technology, Peking University \\
100871 Beijing, China}

\maketitle\abstracts{
A search for a new heavy particle decaying to a pair of vector bosons (WW or WZ) is presented using data from the CMS detector corresponding to an integrated luminosity of $35.9~\mathrm{fb}^{-1}$ collected in proton-proton collisions at a centre-of-mass energy of $13~\mathrm{TeV}$ in 2016. One of the bosons is required to be a W boson decaying to $e\nu$ or $\mu\nu$, while the other boson is required to be reconstructed as a single massive jet with substructure compatible with that of a highly-energetic quark pair from a W or Z boson decay. The search is performed in the resonance mass range between 1.0 and 4.4.  The largest deviation from the background-only hypothesis is observed for a mass near $1.4~\mathrm{TeV}$ and corresponds to a local significance of 2.5 standard deviations. The result is interpreted as an upper bound on the resonance production cross section. Comparing the excluded cross section values and the expectations from theoretical calculations in the bulk graviton and heavy vector triplet models, spin-2 WW resonances with mass smaller than $1.07~\mathrm{TeV}$ and spin-1 WZ resonances lighter than $3.05~\mathrm{TeV}$, respectively, are excluded at 95\% confidence level.~\cite{Sirunyan:2018iff}}

\section{Introduction}

There are several theoretical models that motivate the existence of heavy particles that decay to
pairs of bosons. These models usually aim to answer open questions of the Standard Model (SM) such as the integration of gravity into the SM using extra dimensions.
Popular examples of such models include
the bulk scenario~\cite{Fitzpatrick:2007qr} of
the Randall-Sundrum Warped Extra Dimensions
model~\cite{Randall:1999ee}, and the composite heavy vector triplet (HVT) model~\cite{Pappadopulo:2014qza}.
The composite HVT generalizes a large number of explicit models predicting spin-1 resonances, which can be described by a rather small set of parameters.

In this paper, we describe a search for a heavy resonance decaying to
a pair of vector bosons, one being a W boson decaying to an electron or muon
and a neutrino,
the other being a vector boson decaying to a $q\bar{q}^{(')}$ pair.  The analysis is based on the
proton-proton collision data set collected by the CMS experiment at the LHC in 2016, at a
centre-of-mass energy of $13~\mathrm{TeV}$. The collected data
correspond to an integrated luminosity of $35.9~\mathrm{fb}^{-1}$.

\section{CMS Detector}
The central feature of the CMS apparatus is a superconducting solenoid of 6m internal diameter, providing a magnetic field of 3.8T. Contained within the superconducting solenoid volume are a silicon pixel and strip tracker, a lead tungstate crystal electromagnetic calorimeter (ECAL), and a brass and scintillator hadron calorimeter (HCAL), each composed of a barrel and two endcap sections. Muons are measured in gas-ionization detectors embedded in the steel flux-return yoke outside the solenoid. Extensive forward calorimetry complements the coverage provided by the barrel and endcap detectors. A more detailed description of the CMS detector, together with a definition of the coordinate system used and the relevant kinematic variables, can be found in Ref.~\cite{Chatrchyan:2008aa}.

\section{Signal extraction and Background estimation}

For this analysis, a novel signal extraction method based on a
2D maximum likelihood fit is introduced. 
%Many gains are
%expected by this approach:
%Since the correlations between $m_{WV}$ and $m_{jet}$ are encoded in the fit, larger sideband regions are used decreasing the statistical uncertainty of the background.
%Since the full jet mass range is used, different searches on $WW, WZ, WH$ are encoded in the same analysis by just using the signal shape of the jet mass sideband. More complicated models
%such as the heavy triplet model can be easily investigated since all resonances can also be fitted together.
The main challenge in this signal extraction is due to the very large correlations between the mass of the jet and its transverse momentum which have to be encoded by the fit.

The signal and background yields are determined through
a maximum likelihood fit.
%performed in the
%portion of the ($m_{WV}$, $m_{jet}$) plane defined by the event
%selection described in Section~\ref{sec:selection}.
The fit is performed using 2D templates for signal and
background processes, starting from simulation and introducing shape uncertainties that model the
difference between data and simulation in the full search range.

Two classes of background events are considered:
\begin{enumerate}
    \item A W+jets background, consisting of a lepton and at least one jet arising from a quark or gluon
    mistagged as a $V$ jet. In addition to $W\to l\nu$+jets, this background also includes $t\bar{t}$
    production where the leptonically decaying $W$ boson was reconstructed, but the merged jet
    corresponds to a random combination of jets in the event and not to a $W$ boson or
    a top quark decay.
    \item A $W$+$Vt$ background, peaking in $m_{jet}$ while smoothly falling in $m_{WV}$.
    This background is dominated by $t\bar{t}$ production while sub-dominant
    contributions include SM diboson and single top production.
\end{enumerate}
Each background is modelled by a separate shape pdf based on its properties.

\subsection{Signal}

The probability density function (pdf) of $X \to WV$ events in
the ($m_{WV}$, $m_{jet}$) plane is modelled as:
\begin{equation}
P_\mathrm{sig}(m_{WV},m_{jet}|m_{\mathrm{X}}) = P_{WV}(m_{WV}|m_{\mathrm{X}},\theta_1) \, P_{j}(m_{jet}|m_{\mathrm{X}},\theta_2).
\end{equation}
The pdfs $P_{WV}$ and $P_j$ are represented by double
Crystal Ball~\cite{Oreglia:1980cs,Gaiser:1982yw} functions, and an additional exponential function is used in the jet mass dimension
in LP events to model the tails of the soft-drop jet mass distribution. The parameters of the functions
are described by uncorrelated polynomial interpolations, obtained by fitting the simulated signal sample distributions with the pdfs for different values of the resonance mass $m_{\mathrm{X}}$. The experimental resolution for $m_{jet}$ is around 10\%, and for $m_{WV}$ it ranges from 6\% at $1~\mathrm{TeV}$ to 4\% at $4~\mathrm{TeV}$.

%For the signal, the dependence of the shape parameters on the
%esonance mass is found to depend on the nature of the $V$ jet (e.g.
%$W$ or $Z$) and the lepton flavour. The signal yields in the different
%signal categories are expressed as a function of the integrated
%luminosity of the sample and the product of the signal acceptance and efficiency, treated as nuisance
%parameters, so that the resonance production cross section is
%determined in a combined fit to data in the four categories.

\subsection{Non-resonant background: W+jets}

The $W$+jets background shape is described as a conditional probability
of $m_{WV}$ as a function of $m_{jet}$:
\begin{equation}
P_{W+jets}(m_{WV},m_{jet}) = P_{WV}(m_{WV}|m_{jet},\theta_1) \, P_{j}(m_{jet}|\theta_2).
\end{equation}
The conditional probability is essential to take into account the large
correlations between $m_{jet}$ and $m_{WV}$. Those correlations arise from the
strong dependence of the jet mass on the jet $p_{T}$ during the hadronization
process.
The 2D conditional templates, $P_{WV}$, are constructed from simulated events, starting before the detector simulation stage.
For each event in the background samples, jets are clustered from stable particles
using the same substructure algorithms as during event reconstruction. Consequently, a scale and resolution model is derived for both $m_{jet}$ and
$m_{WV}$ as a function of generated jet $p_{T}$ by comparing the reconstructed and generated variables. Smooth templates are then populated as sums of 2D Gaussian distributions, where
the mean values of the Gaussians correspond to the true value of $m_{jet}$ and $m_{WV}$, shifted by the derived scale model, and the 2D covariance matrix
is given by the resolution model.
This technique is similar to the kernel-estimation procedure given in Ref.~\cite{Cranmer:2000du} but uses the simulation and the exact resolution model instead of starting from reconstructed events.
The final step is to smooth the tails for high values of $m_{WV}$ ensuring there are no empty bins in the templates.
The smoothing is performed by fitting events in each $m_{jet}$ bin with $m_{WV} >2.0$ using an exponential
function and then using the function values to populate the tails for $m_{WV}>2.5~\mathrm{TeV}$.
The $P_{j}$ shapes are one-dimensional (1D) histograms derived directly from reconstructed simulated events, in contrast to the $P_{WV}$ shapes discussed above.

%
%For both the $P_{j}$ and $P_{WV}$ components, nuisance parameters are introduced to account for %differences between data and simulation.
%The most important difference is attributed to the different $p_{T}$ spectrum of the jets in the simulation. %The template
%construction is repeated by adding event weights corresponding to a harder (softer) spectrum, and the %pdf is interpolated between these alternative templates. An additional uncertainty lies in the choice of %the scale/resolution model, which is estimated by varying the scale as functions of $m_{jet}$ and %$m_{WV}$.
%The derived shapes are found to be in agreement with the simulated events, validating the template %construction procedure.
%This procedure implicitly assumes that a single component can account for the sum of the $t\bar{t}$ %events with an arbitrary fraction of reconstructed $W$+$V$ jet and W+jets
%contributions. A variation of the relative fractions is found to
%translate into a change in the average $p_{T}$ spectrum, which is taken into account as a systematic %uncertainty.
%

\begin{figure}[htbp]
    \centering
        \includegraphics[width=0.4\textwidth]{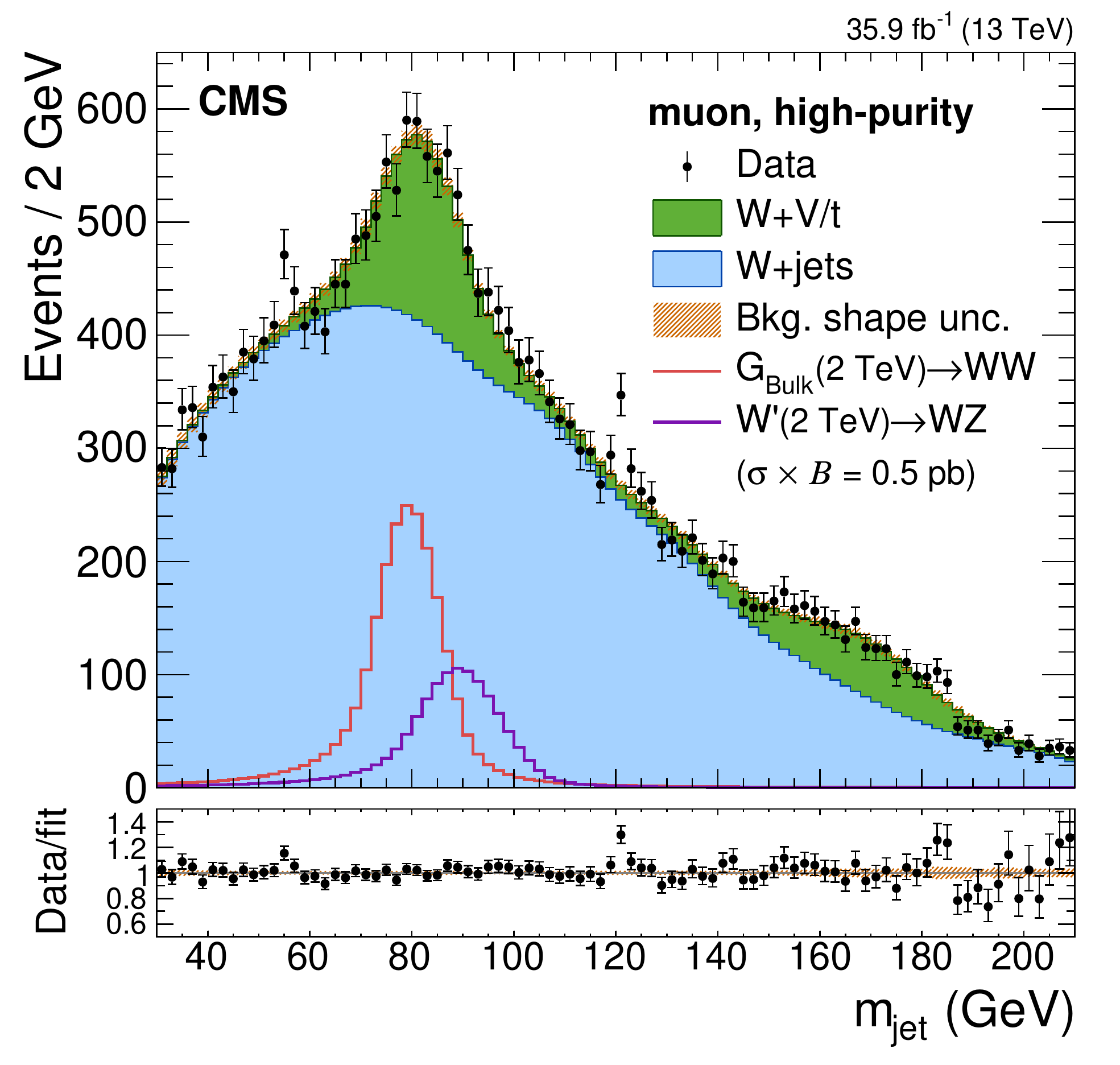}
        \includegraphics[width=0.4\textwidth]{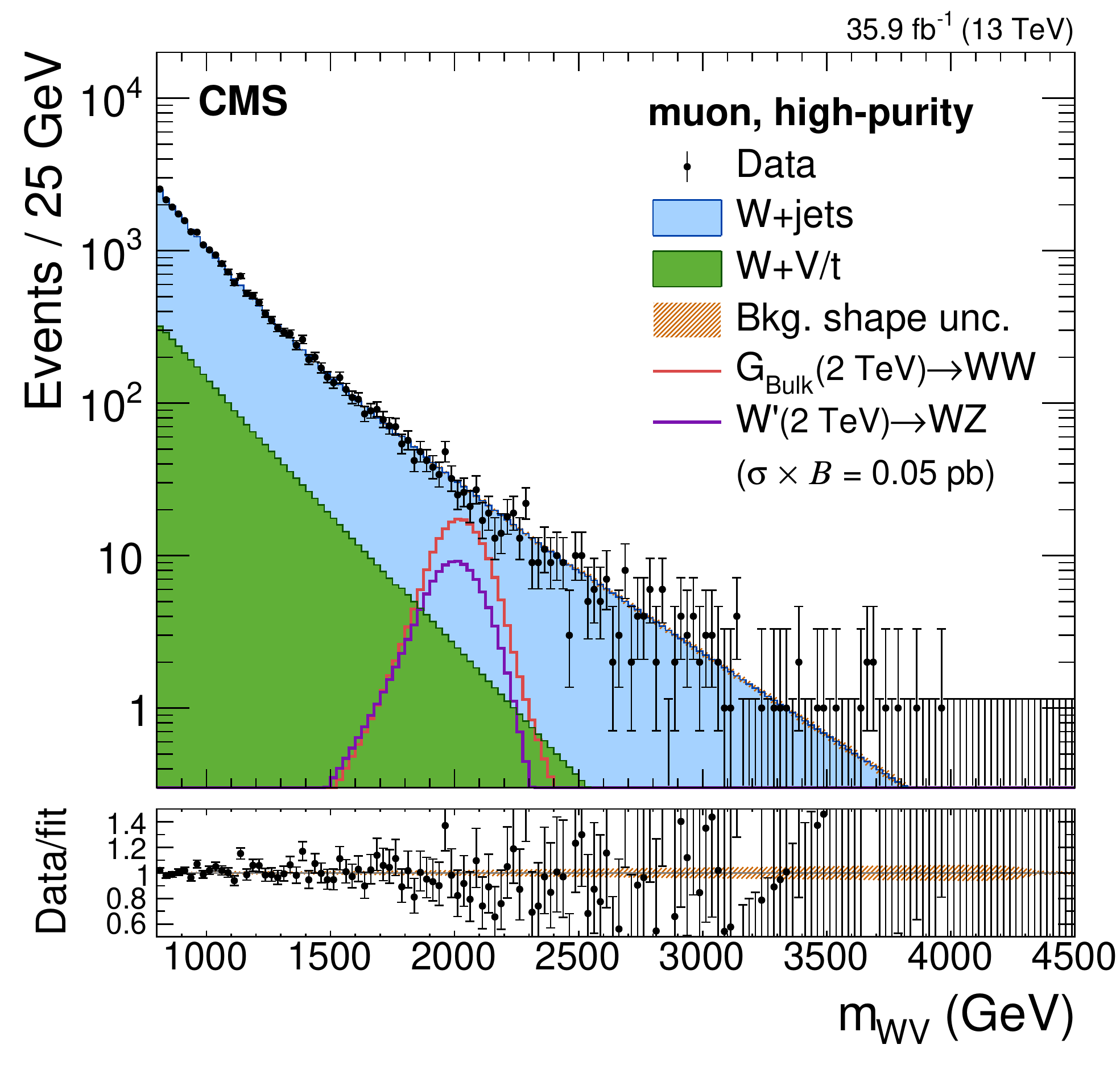}
    \caption{Comparison between the fit result and data distributions of $m_{jet}$ (left) and $m_{WV}$ (right)
        in the muon HP category.
        The background shape uncertainty is shown as a shaded band, and the statistical uncertainties of the data are shown as vertical bars.
        No events are observed with $m_{WV}>4.5~\mathrm{TeV}$.
        Example signal distributions are overlaid, using an arbitrary
        normalization that is different in the upper and lower plots.
        \label{fig:HP}}

\end{figure}

\subsection{Resonant background: W+V}

The $W$+$Vt$ background is modelled as :
\begin{equation}
P_{W\mathrm{+}Vt}(m_{WV},m_{jet}) = P_{WV}(m_{WV}|\theta_1) \, P_{j}(m_{jet}|m_{WV},\theta_2).
\end{equation}
In this case, $P_{WV}$ is a 1D template constructed in the same way as for the $W$+jets background, and the smoothing of its tail with an exponential function is performed for $m_{WV} > 1.2~\mathrm{TeV}$.
$P_{j}$ is described by two peaks: a peak around the $W$ boson mass dominated by top quark events
where only the $W \to q\bar{q}'$ was reconstructed inside the large-radius jet, and a peak
around the top quark mass where the $W$ boson and the b quark decays are merged.
These peaks are modelled by two double Crystal Ball plus one exponential function, whose parameters are described by uncorrelated polynomial functions of $m_{WV}$.
The presence of both jet peaks allows additional scrutiny, since the relative fraction
of the two peaks as a function of the resonance mass provides a robust validation of the top quark $p_{T}$ spectrum convolved
with effects from jet grooming. Different shapes are used in the individual event categories to
account for differences in the event kinematic distributions.

\section{Results and Summary}
\begin{figure}[h]
    \centering
        \includegraphics[width=0.9\textwidth]{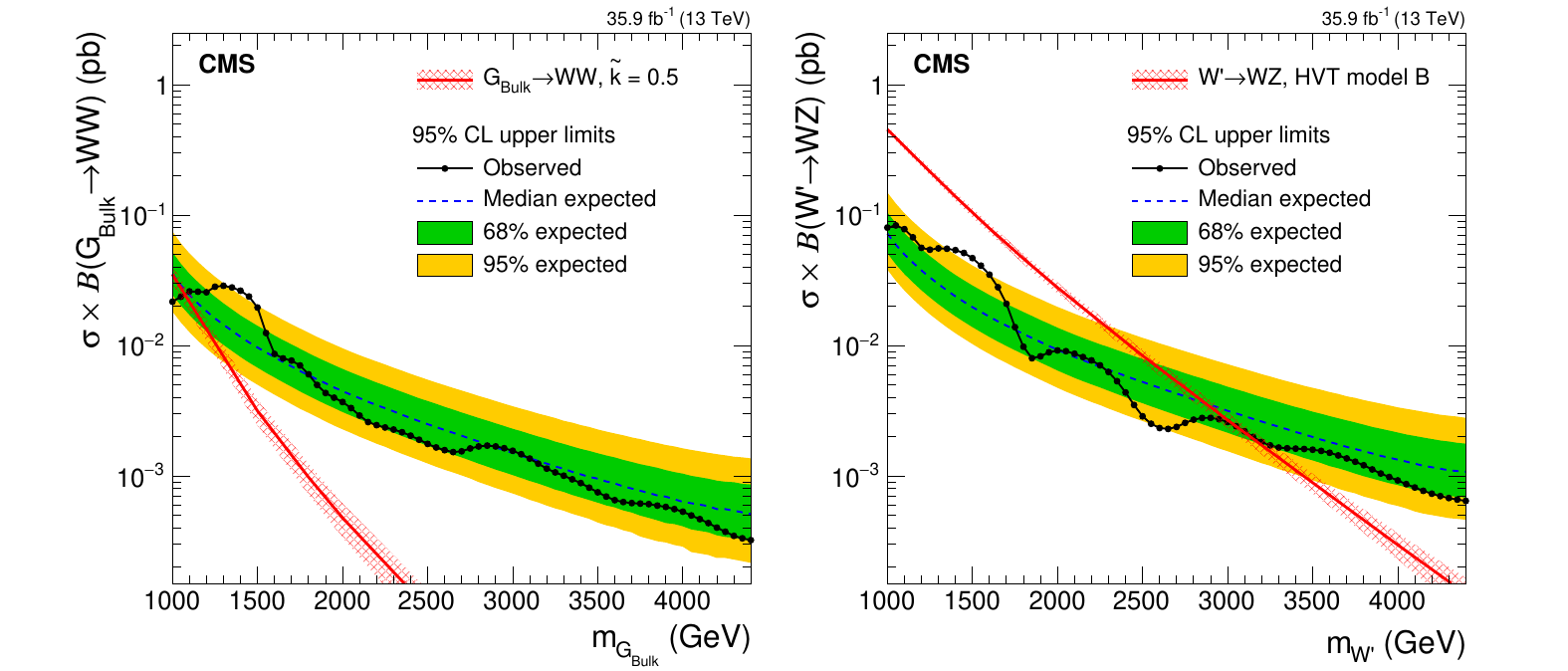}
    \caption{Exclusion limits on the product of the production cross section and the branching fraction for a new spin-2 resonance decaying to $WW$
        (left) and for a new spin-1 resonance decaying to $WZ$ (right), as a
        function of the resonance mass
        hypothesis. Signal cross section uncertainties are shown as red cross-hatched bands.\label{fig:exclusion_limits}}
\end{figure}

The results are interpreted in terms of exclusion limits for the
benchmark signal models of BulkGraviton and $W'$.
We provide model-independent limits, which are not coupled to the relative normalizations of the benchmark models.
We expect any model-dependent effects on the acceptance and selection efficiency to be covered by the PDF and scale uncertainties.
Figure~\ref{fig:exclusion_limits} shows the upper exclusion limits on the product of the resonance production cross section
and the branching fraction to $WW$ or $WZ$ as a function of the resonance mass.

%The observed limits for the $WW$ signal range from 29 fb at $1.3~\mathrm{TeV}$ to 0.32 fb at $4.4~\mathrm{TeV}$,
%while for the $WZ$ signal they range from 84 fb at $1.05~\mathrm{TeV}$ to 0.64 fb at $4.4~\mathrm{TeV}$.
By comparing these limits to the expected cross sections from the benchmark theoretical models,
$WW$ resonances lighter than $1.07~\mathrm{TeV}$ and $WZ$ resonances lighter than $3.05~\mathrm{TeV}$  are excluded at 95\% confidence
level (CL), using the asymptotic approximation
%~\cite{Cowan:2010js} 
of the
$CL_{s}$\ method
%~\cite{Junk:1999kv,Read:2002hq}
.

\section*{Acknowledgments}

I would like to thank the Moriond - EW 2018 organizers for their hospitality and
the wonderful working environment. I acknowledge the support from
the National Natural Science Foundation of China, under Grants No.11661141008.

\end{document}